\documentclass[useAMS,usenatbib]{mnras}
\usepackage{natbib}
\usepackage{graphicx}

\title[Galactic Mass Function]{Was the Milky Way a chain galaxy? Using the IGIMF theory to constrain the thin-disk star formation history and mass}

\author[Zonoozi et al.]
{Akram Hasani Zonoozi$^{1}$\thanks{
E-mail:  \mbox{a.hasani@iasbs.ac.ir} (AHZ)},
Hamidreza Mahani$^{1}$,  Pavel Kroupa$^{2, 3}$,
\\
$^{1}$Department of Physics, Institute for Advanced Studies in Basic Sciences (IASBS), PO Box 11365-9161, Zanjan, Iran\\
$^{2}$Helmholtz-Institut f\"ur Strahlen-und Kernphysik (HISKP), Universit\"at Bonn, Nussallee 14-16, D-53115 Bonn, Germany\\
$^{3}$Charles University in Prague, Faculty of Mathematics and Physics, Astronomical Institute, V Hole\v{s}ovi\v{c}k\'ach 2, CZ-180 00 \\
Praha 8, Czech Republic\\}

\begin{document}

\date{Accepted .... Received}

\pagerange{\pageref{firstpage}--\pageref{lastpage}} \pubyear{2013}

\maketitle

\label{firstpage}

\maketitle

\begin{abstract}

The observed present-day stellar mass function (PDMF) of the solar neighborhood is a mixture of stellar populations born in star-forming events that occurred over the life-time of the thin disk of the Galaxy. Assuming stars form in embedded clusters which have stellar initial mass functions (IMFs) which depend on the metallicity and density of the star-forming gas clumps, the integrated galaxy-wide IMF (IGIMF) can be calculated. The shape of the IGIMF thus depends on the SFR and metallicity. Here, the shape of the PDMF for stars more massive than $1\,M_\odot$ in combination with the mass density in low-mass stars is used to constrain the current star-formation rate (SFR), the star formation history (SFH) and the current stellar plus remnant mass ($M_*$) in the Galactic thin disk. This yields the current SFR, $\dot{M}_*= 4.1^{+3.1}_{-2.8}~M_\odot$yr$^{-1}$, a declining SFH and $M_*=2.1^{+3.0}_{-1.5}\times 10^{11}M_\odot$, respectively, with a V-band stellar mass-to-light ratio of $M_*/L_V=2.79^{+0.48}_{-0.38}$.These values are consistent with independent measurements. We also quantify the surface density of black holes and neutron stars in the Galactic thin disk. The invariant canonical IMF can reproduce the PDMF of the Galaxy as well as the IGIMF, but in the universal IMF framework it is not possible to constrain any of the above Galactic properties. Assuming the IGMF theory is the correct framework and in combination with the vertical velocity dispersion data of stars, it follows that the Milky Way would have appeared as a chain galaxy at high redshift.

\end{abstract}


\section{INTODUCTION}\label{Sec:INTRO}

The initial distribution of stellar masses that form together in one star formation event, the so-called "initial mass function" (IMF), plays an important role in  astrophysics. The IMF is needed in order to determine the chemical enrichment and population synthesis modeling of galaxies \Citep{Tinsley80}, and to estimate their mass-to-light ratios and baryonic mass content from the observed luminosities. It is also an important key in understanding the formation and  dynamical evolution of star clusters and galaxy evolution.

The direct observational determination of the IMF is a very difficult task, as it is based on star counts and remnant-star corrections.  Much work has been done to constrain the shape of the IMF on star-cluster and galaxy scales. Many biases, assumptions and systematic uncertainties are involved in the derivation of the IMF, i.e., measuring the luminosity function from the observational data, converting it to the PDMF using a mass-luminosity relation, and correcting the PDMF for SFH, stellar evolution, binary fraction and dynamical evolution of star clusters (for reviews see \Citealt{Kroupa02a, Chabrier03, Bastian10, Kroupa13, Offner14}). 

The universality or environmental dependency of the IMF is still under debate. From empirical results, the IMF appears to be largely invariant in individual star forming events within the present-day Local Group  \Citep{ Bastian10}. But theoretically, both the Jeans mass and the self-regulation arguments predict the IMF to be top-heavy under low metallicity and high-temperature star-forming conditions \Citep{Kroupa13}. Although there is no clear evidence for significant variations of the slope of the IMF in local star clusters \Citep{Kroupa02a, Bastian10}, there is increasing observational and theoretical evidence that the IMF shows variations in extreme environments \citep{ Dabringhausen12, Marks12}.
Notably, the low-metallicity star burst regions 30 Doradus in the LMC \citep{Schneider18, Banerjee2012} and the massive cluster NGC 796 in the LMC/SMC bridge region \citep{Kalari18} have been found to have top-heavy IMFs.

Concerning galaxy scales, extragalactic observational data show a systematic variation of the galaxy-wide IMF. The IMF appears to be flatter (more top-heavy) for active galaxies with high star formation rates (SFRs) \Citep{ Hoversten08, Lee09, Meurer09, Habergham10, Gunawardhana11}, while dwarf galaxies and low surface brightness galaxies appear to have top-light galaxy-wide IMFs \Citep{Ubeda07, Lee09, Watts18}. There are also competing models for the deficiency of high mass stars in low density regions based on the stochasticity of a universal IMF \citep{Koda12} in a study of deep $H\alpha$ observations of the outskirts of M83. Recently, \Citet{Zhang18} found evidence for a top-heavy  stellar IMF in a sample of starburst galaxies at redshifts of approximately two to  three. 
A varying IMF at the  low mass end has been found in early type galaxies \Citep{, vanDokkum11, Spiniello15, Conroy17} and ultra faint dwarf galaxies that appear to be more bottom-heavy for more massive galaxies and more bottom-light with  decreasing galactic mass and metallicity, respectively  \Citep{Geha13, Gennaro18}.

In order to address the problem how the galaxy-wide IMF may be related to the IMF in star forming regions \citet{Kroupa03} formulated the integrated galaxy IMF (IGIMF) theory. Assuming that all stars form in embedded clusters \Citep{Lada03, Kroupa05, Megeath16} the galaxy-wide IMF follows from adding the IMFs of all freshly formed embedded clusters. These range from small masses near $5\,M_\odot$ comparable to the little clusters observed in e.g. the Taurus-Auriga star-forming cloud \Citep{Joncour18}, up to the most massive cluster allowed by the current SFR. Based on this theory, the galaxy wide IMF may significantly deviate from the star-cluster-scale IMF. The IGIMF theory implies the IMF in massive galaxies with high SFRs to be top-heavy while in low-mass galaxies which have a low-level of star formation activity it becomes top-light \citep{Yan2017}. A full IGIMF grid in dependence of the SFR and metallicity of a galaxy is provided by \citet{Jerabkova18}.

Using the IGIMF theory, the chemical evolution of the solar neighborhood \Citep{Calura10},  of star-forming \citep{Kopen2007} and of elliptical galaxies \Citep{Recchi09, Fontanot2017} has been studied. Many other aspects of galaxy evolution have been shown to be resolved using the IGIMF theory (for a review see \citealt{Pflamm11}), and for example the radial $H\alpha$ cutoff in disk galaxies with UV extended disks is a normal consequence of the IGIMF theory \citep{Pflamm08}.

Indeed, because the shape of the observationally constructed stellar IMF in the Galactic field for stars with $m>1\,M_\odot$  appears to be steeper than the stellar IMF deduced from individual star-forming regions \citep{Scalo86, Rybizki15, Yan2017}, we here address this difference in shape by applying the IGIMF theory with a time-dependent SFR to constrain the star-formation history of the Galaxy and thus its present-day SFR.  In this paper, by studying the present-day mass function (PDMF) of the MW, the metallicity- and SFR-dependent IGIMF is being tested and compared with the invariant canonical IMF. In Sec. 2, the set-up of our models is described. We  compare our models with the observed PDMF of the MW to constrain the properties of  the Galactic thin disk in Sec. 3, and give our conclusions in Sec. 4.

\begin{table*}
 \centering
 \caption{The here-used model-grid of $M_{tot}$, $\tau$ and corresponding $b$ values. For example, the increasing-with-time SFR with $\tau= -2.20$Gyr, corresponds to $b=4.59$, that is, the present-day SFR is larger than the average by a factor of 4.59 (Eq. \ref{b}).}
 \begin{tabular}{ccccccccccccccccc}
 \hline
 \hline
 $M_{tot} [M_{\odot}]$  & $ 10^9$ & $ 10^{10}$& $10^{11}$& $2\times 10^{11}$& $4\times 10^{11}$& $10^{12}$& $10^{13}$& $ 10^{14}$\\
 \hline
 \hline
 $\tau [Gyr]$  & $1.55$ & $2.20$ & $2.80$ & $3.25$ & $3.80$ & $4.90$ & $6.20$ & $8.00$ & $10.50$ & $15.00$ & $23.00$ & $50.00$ & $\infty $ \\
          & $-50.00$ & $-23.00$ & $-15.00$ & $-10.50$ & $-8.00$ & $-6.20$ & $-4.90$ & $-3.80$ & $-3.25$ & $-2.80$ & $-2.20$ & $-1.55$ & $-1.20$ & $-1.00$ \\
 \hline
 $b$  & $0.01$ & $0.05$ & $0.10$ & $0.15$ & $0.20$ & $0.30$ & $0.40$ & $0.50$ & $0.60$ & $0.70$ & $0.80$ & $0.90$ & $1.00$\\
       &  $1.10$ & $1.23$ & $1.37$ & $1.55$ & $1.75$ & $2.01$ & $2.35$ & $2.84$ & $3.23$ & $3.67$ & $4.59$ & $6.46$ & $8.33$ & $10.00$\\
 \hline
 \hline
 
 \end{tabular}
 \label{tab:orbits}
 \end{table*}

\section{method and model ingredients}\label{Sec:Results}

\subsection{The stellar IMF}

We distinguish between the galaxy-wide IMF, $\xi_{gal}$, and the stellar IMF, $\xi$, found in star-forming units, i.e., in embedded clusters. Our calculations are based on \emph{(i)} assuming $\xi_{gal}=\xi$  which is an invariant IMF with the Salpeter-Massey power-law slope $\alpha_3=2.3$ for stars more massive than $1M_\odot$, and \emph{(ii)}  that $\xi_{gal}=$ IGIMF which varies with the SFR of a galaxy.  The invariant IMF can be well described by the canonical IMF, a two-part power-law function \citep{Kroupa01, Kroupa13}. The number of stars in the mass interval $m$ to $m+dm$ is $dN=\xi(m)dm$, where

\begin{eqnarray}
\xi(m\leq m_{max})=k\, 
\left\{
      \begin{array}{ll}
2\, m^{-\alpha_1}   & 0.08M_\odot\leq m < 0.5M_\odot,  \\
\,\,\,\, m^{-\alpha_2}   & 0.5M_\odot\leq m < 1M_\odot,  \\
\,\,\,\, m^{-\alpha_3}    &   1M_\odot\leq m \leq m_{max}(M_{ecl}),
       \end{array}
        \right. \label{MF}
\end{eqnarray}
where, $\alpha_1=1.3$, $\alpha_2=2.3$ and $k$ is the normalization constant. For the invariant canonical IMF, $\alpha_2=\alpha_3=2.3$ is the constant Salpeter-Massey slope. The function $m_{max}=WK_1 (M_{ecl})$ is the $m_{max}-M_{ecl} $  relation between the mass of the most massive star in the cluster and the stellar mass of the cluster \citep{Weidner06, Weidner10, Weidner13}. More recent observational surveys support the existence of this relation \Citep{ Ramirez16,  Stephens17, Oh18} which is most probably the result of self-regulated growth during the formation of the embedded cluster \citep{Kroupa13}.\Citet{Andrews13, Andrews14} claim to falsify the $m_{max}-M_{ecl}$ relation of \Citet{Weidner10} using  the data obtained from the young starburst dwarf galaxy, NGC 4214, and starbursting spiral galaxy, M83 .
Because the $H_\alpha$ luminosities of the unresolved clusters in these galaxies are above their expected value assuming $m_{max}-M_{ecl}$ relation as a truncation limit with assumed random sampling of stellar masses, they argue that this relation is not correct. \Citet{Weidner14} show instead that the data of \Citet{Andrews13} are well consistent with the $m_{max} = WK_1 (M_{ecl})$ relation, because the random sampling procedure adopted by \Citet{Andrews13} is not correct as it implies the actual average $m_{max}$ to be significantly below the truncation and furthermore  the ages and flux measurements readily make the observations well consistent with the relation.

 \subsection{The galaxy wide IMF}

Assuming that the star formation process takes place in embedded star clusters, the galaxy-wide IMF is defined as an integral over the embedded cluster IMF, $\xi(m)$, weighted with the embedded star cluster MF, $\xi_{ecl}(M) $,

\begin{eqnarray}
\xi_{IGIMF}(m,\psi(t))=\,\,\,\,\,\,\,\,\,\,\,\,\,\,\,\,\,\,\,\,\,\,\,\,\,\,\,\,\,\,\,\,\,  \nonumber\\
\int_{M^{min}_{ecl}}^{M^{max}_{ecl}(\psi(t))}\xi(m\leq m_{max})\xi_{ecl} (M_{ecl},\psi(t)) dM_{ecl} ,  \label{IGIMF}
\end{eqnarray}
where the IMF depends on $M_{ecl}$ and $[Fe/H]$, and  $M_{ecl, max}$ depends on star formation rate,$\psi(t)$.

 For a sample of ultra-compact dwarf galaxies (UCDs) and  MW globular clusters the data suggest that the high-mass IMF is more top-heavy (flatter) in more massive, denser and metal-poorer environments \citep{ Dabringhausen09, Dabringhausen12, Marks12}. This may partially be a result of cosmic ray heating \citep{Pap11} and cloud-core coagulation in very dense embedded  forming proto-clusters \citep{Dib07}.
 However, an increased X-ray luminosity was not found in another sample of UCDs \Citet{Pandya16}. The difference in the two samples could be possibly based on different formation scenarios of UCDs, i.e., the single monolithic collapse \Citep{Murray09, Dabringhausen09} or the merging of cluster complexes \citep{Kroupa98, Fellhauer02a, Fellhauer02b, Bruns11}. According to \Citet{Marks12} and \Citet{Dabringhausen12}, that is biased to only compact UCDs, a top-heavy IMF is expected for UCDs that form monolithically while it could be canonical or even top-light for UCDs that  form from the merging of many star clusters.

 Due to the correlation between the metallicity and molecular cloud core density ($\rho_{cl}$), $\alpha_3$ depends on both metallicity and density of the embedded-cluster forming molecular cloud cores \Citep{Marks12}:

\begin{eqnarray}
\alpha_3=
\left\{
       \begin{array}{ll}
          2.3    & x<-0.87,  \\
          -0.41x+1.94     & x>-0.87,
 \end{array}
          \right. \label{alpha3}
\end{eqnarray}
where $x=-0.14[Fe/H]+0.99\log_{10}(\rho_{cl}/10^6M_\odot pc^{-3} )$. A metallicity and density dependent top-heavy IMF is successfully used to explain the observed mass-to-light $(M/L)$ ratios of star clusters in M31 which show an inverse trend with metallicity \Citep{Zonoozi16, Haghi17}.

 
The  NGC 346 region in the SMC appears to have a Salpeter slope of the PDMF \Citep{Sabbi08} which is in  agreement with the above equation adopting it's metallicity, $Z=0.2 Z_\odot$ \Citep{Bouret03}  and density,  $\rho=10^3 M_\odot/pc^2$ \Citep{Sabbi07}.

The embedded-cluster-forming cloud core density, $\rho_{cl}=3M_{cl}/4\pi r_h^3$, where $M_{cl}$ is the original cloud core mass in gas and stars which is three times the stellar mass of the embedded cluster for a star formation efficiency of 33\% \citep{Lada03, Megeath16}. The initial half-mass radius of the cloud core, $r_h$, follows from an analysis of the initial conditions of star clusters using the observed distribution of binding energies of binaries by \citet{Marks12a},

\begin{equation}
r_h (pc) = 0.1\times\left(\frac{M_{cl}}{M_{\odot}}\right)^{0.13}. \label{MKrelation}
\end{equation}

The maximum stellar mass $m_{max}$ in Eq. 2  is  the function $m_{max}= WK_1(M_{ecl})$ in Eq. \ref{MF}. Assuming the upper-most possible stellar mass is 150$M_{\odot}$ \citep{Weidner04,Figer05,Oey05,Koen06,Maiz07,Banerjee2012}, the normalization constant $k$ and $m_{max}$ are determined by solving the following equations:
\begin{eqnarray}
M_{ecl}=\int_{0.08\,M_\odot}^{m_{max}} m\,\,\xi(m)dm, \label{Mecl}
\end{eqnarray}

\begin{eqnarray}
1=\int_{m_{max}} ^{150\, M_\odot} \xi(m) dm. \label{Meclnorm}
\end{eqnarray}

The embedded cluster mass function, $\xi_{ecl}(M_{ecl})$,  is assumed to be a power law function,  becoming top-heavy in galaxies with a high SFR \citep{Weidner13b, Zhang99, Recchi09, Yan2017},
\begin{eqnarray}
\xi_{ecl}(M_{ecl})=K_{ecl} M_{ecl}^{-\beta}, \nonumber\\
\beta=-0.106 \log_{10} \psi(t) + 2, \label{beta}
\end{eqnarray}
where the number of embedded clusters with stellar masses in the interval $M_{ecl}$ and $M_{ecl}+dM_{ecl}$ is $dN_{ecl}=\xi_{ecl}(M_{ecl})dM_{ecl}$.

The lower limit for an embedded cluster mass is assumed to be $M_{ecl}^{min}=5 M_\odot$, corresponding to the smallest star-forming stellar cluster known \Citep{Kirk12, KroupaBovier03, Joncour18}. The mass of the most massive embedded cluster, $ M_{ecl}^{max}= WK_2(SFR)$, and the normalization constant $K_{ecl}$ in Eq. \ref{beta} are determined by simultaneously solving the following equations:

\begin{eqnarray}
M_{tot-10~Myr}=\int_{M_{ecl}^{min}}^{M_{ecl}^{max}} M_{ecl}\xi_{ecl}(M_{ecl})dM_{ecl}, \nonumber\\
1=\int_{M_{ecl}^{max}} ^{10^9 M_\odot} \xi_{ecl}(M_{ecl}) dM_{ecl}, \label{Mtot10}
\end{eqnarray}
where $M_{tot-10~Myr}= \psi(t)\times\delta t $ is the total stellar mass formed within $\delta t=10$ Myr assuming the galaxy-wide SFR, $\psi(t)$, remains constant over the time $\delta t$. The upper integration limit of $10^9 M_\odot$ is adopted as a physical limit of embedded cluster masses \citep{Dab08}.  Here, $\delta t$ is the time scale over which the interstellar medium of the galaxy forms a new population of embedded clusters which optimally fill the embedded cluster mass function $\xi_{ecl}(M_{ecl})$ (see \citealt{Kroupa13, Schulz15} and \citealt{Yan2017} for a discussion and additional references).  A Fortran code \emph{GWIMF} is available to calculate the IGIMF as a function of SFR and metallicity \footnote{ https://github.com/ahzonoozi/GWIMF}. See also \citet{Yan2017} for a python module for the same purpose.

 \subsection{The present day stellar mass function}

To constrain the physical properties of the MW, its total stellar mass, stellar mass-to-light ratio, and SFR, we calculate the Galactic PDMF under different conditions. The PDMF of the Galaxy refers to the MF of all main sequence stars which can be observed today in the Galactic disc in the vicinity of the Sun. It can be determined by converting the observed luminosity function of the Galactic-field stars to a mass distribution adopting a mass-luminosity relation and correcting the star counts for stellar evolution  \citep{Salpeter55,Miller79,Scalo86, Kroupa93, Rybizki15,Mor17}. We assume in this analysis that the shape of the observationally constrained PDMF to be representative for the whole Galactic disc.

The PDMF of the Galaxy is determined principally by the IMF and the star formation history. 
Combining these ingredients with stellar evolution one can calculate the PDMF and the time-evolution of the integrated light of the system, the mass and mass-to-light ratio via stellar population synthesis (SPS). The PDMF of a galaxy at time $T_G$, that is the number of main sequence stars per interval mass, $\xi_{PDMF}(m,t)$, is
\begin{equation}
\xi_{PDMF}(m_{in},T_G)=\int_0^{T_G} \xi_{gal}(m_{in}<m_{in}^u(t),T_G-t) \psi(T_G-t) dt, \label{pdmf}
\end{equation}
where $\xi_{gal}(m_{in}<m_{in}^u(t),t)$  is the galaxy-wide stellar IMF in units of the number of stars per unit mass interval, with $m_{in}$ being the initial mass, and $m_{in}^u(t)$ is the upper live-star-mass cutoff at time $t$, as determined by stellar evolution. The IMF is normalized such that $ \int_{0.08}^{150} m~\xi_{PDMF}(m)dm=1M_\odot $.  $\psi(t)$ is the SFR in units of $M_\odot yr^{-1}$ . The galaxy-wide PDMF is a time dependent function not only due to the upper mass limit of the live stars but also because of its possible intrinsic dependency on the SFR and metallicity which could vary with time.

We adopt a simple exponential model for the SFH in units of $M_\odot yr^{-1}$ which is characterized by the e-folding timescale $\tau$,

\begin{equation}
 \psi(t)= C \exp ^{(-t/\tau)}, \label{sfr}
\end{equation}
where C is the normalization parameter so that the stellar mass produced over the age of the disk is equal to the total stellar mass of the Galactic disk,

\begin{equation}
 M_{tot}= \int_0^{T_G} \psi(t)dt, \label{mtot}
\end{equation}
where $t$ is the time measured from 10 Gyr ago.  We model both an exponentially increasing and decreasing SFR with negative and positive values of $\tau$, respectively. The age of the MW thin disk is assumed to be $T_G=10$ Gyr in all models \citep{Carraro00}. The different values of the birth parameter, b, which is defined as the ratio of  the current to the average past SFR \Citep{Kennicut94},

\begin{equation}
b=\psi(T_G)/\textless \psi\textgreater, \label{b}
\end{equation}
is given in Table 1.

\begin{table*}
\centering
\caption{Details of the best-fitting parameters obtained for the Milky Way thin disk based on different assumptions for the IMF.
Columns 2 and 3 give the best-fitting values of Galactic initial parameters, the e-folding time scale and total mass converted into MW-thin-disc stars.   Column 4 gives the corresponding $\chi^2$ (Eq. \ref{chi2}).  The best-fitting present-day stellar mass-to-light ratio (including remnants), present-day stellar mass, and present-day SFR are given in column 5, 6 and 7, respectively. The last two columns give the number of black holes and neutron stars in the Galactic thin disk for the best-fitting models.}

\begin{tabular}{ccccccccccccccccccc}
\hline
IMF& $\tau $ & $M_{tot}$ & $\chi^2$ & $M_\ast/L_V$ & $M_\ast$ & $\psi_{t=T_G}$ &$N_{BH}$ &$N_{NS}$ \\
& $[Gyr]$ & $[10^{11}M_\odot]$ & $  $ & $[ M_{\odot}/L_\odot]$ & $[10^{11}M_\odot]$ & $[M_\odot yr^{-1}]$&$[10^{8}]$& $[10^{8}]$\\
\hline
$Canonical $ & $2.8^{+0.5}_{-0.1} $ & $1.0 $ & $ 1.16  $ & $ 1.35 ^{+0.06}_{-0.17}  $ & $0.6\pm{0.01}  $ & $ 1.03^{+0.52}_{-0.08}  $ &1.0& 9.0 \\
\hline
$IGIMF$ &$2.8 ^{+1.7} _{-0.1} $  & $4.0^{+8.6}_{-3.5}  $ & $0.88   $ & $2.79^{+0.48}_{-0.38}$ & $2.1^{+3.0}_{-1.5}$ & $4.1^{+3.1}_{-2.8}$ & 9.8& 47.0\\
\hline
$IGIMF_{\beta1}$ &$2.8 ^{+1.7} _{-0.1} $  & $0.8^{+11.8}_{-0.3}  $ & $0.74   $ & $2.81^{+0.46}_{-0.40}$ & $0.45 ^{+4.6}_{-0.01}$ & $0.82^{+6.92}_{-0.1}$ & $1.51$ & 8.63\\
\hline
$IGIMF_{\beta2}$ &$3.8 ^{+0.7} _{-1.1} $  & $6.0^{+6.6}_{-5.5}  $ & $1.29  $ & $1.38^{+1.9}_{-0.1}$ & $3.7 ^{+1.4}_{-3.1}$ & $12.2^{+0.1}_{-10.9}$ & $8.68$ & 52.10\\
\hline
$ constrained $    $ SFH$ &$ $  & $  $ & $  $ & $$ & $$ & $$ & $$ & \\
\hline
$Canonical$ &$4.9 ^{+0.7} _{-1.1} $  & $1.0  $ & $9.46  $ & $0.87^{+1.15}_{-0.32}$ & $0.6 ^{+0.01}_{-0.01}$ & $3.05^{+0.02}_{-0.62}$ & $1.13$ & 8.81\\
\hline
$IGIMF$ &$4.9 ^{+0.7} _{-1.1} $  & $0.50^{+11.6}_{-0.08}  $ & $3.92  $ & $1.87^{+0.85}_{-0.03}$ & $0.31 ^{+1.07}_{-0.01}$ & $1.53^{+5.63}_{-0.27}$ & $0.59$ & 4.35\\
\hline
\end{tabular}
\label{tab:orbits}
\end{table*}

\begin{figure*}
\begin{center}
\includegraphics[width=175mm]{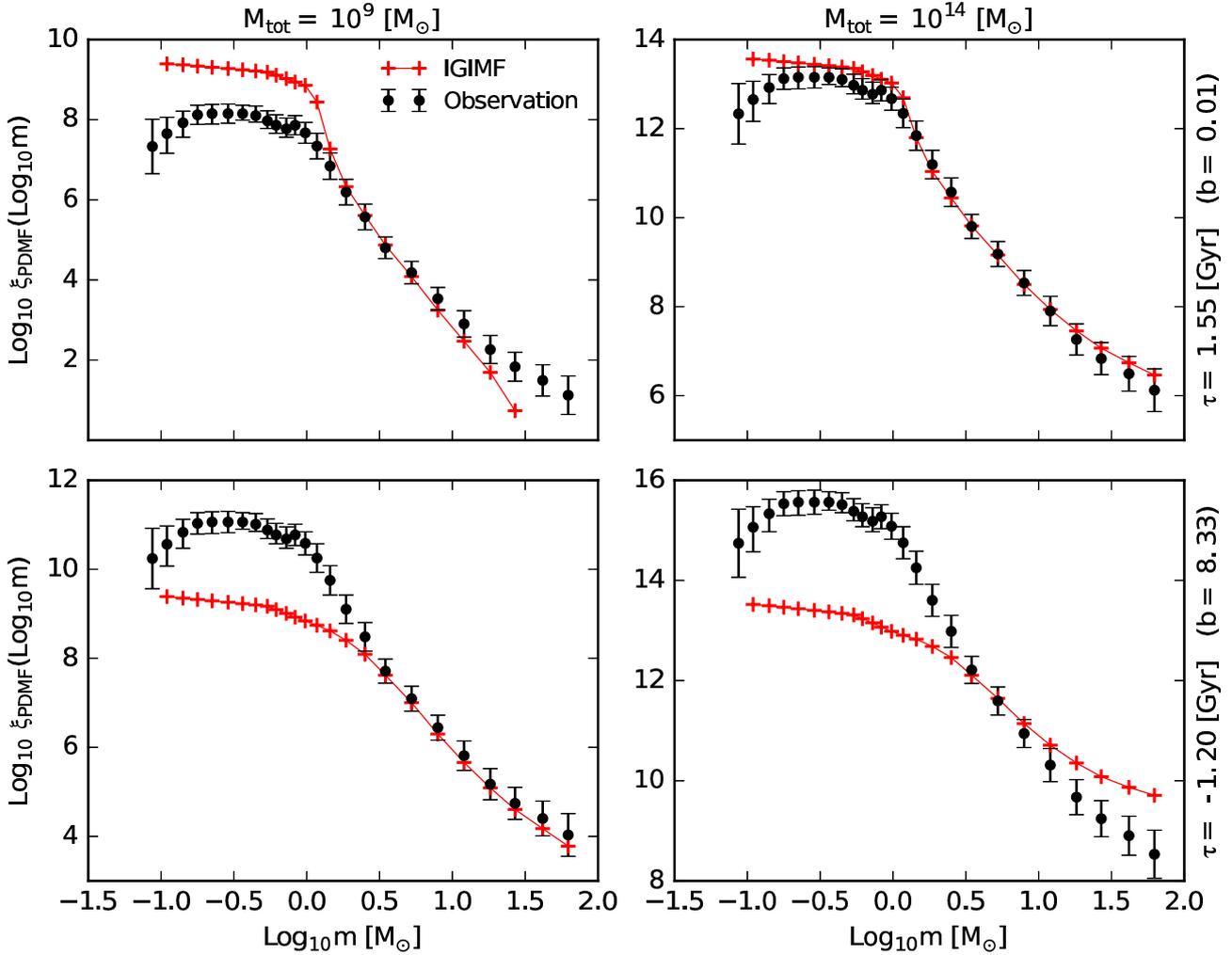}
\caption{ The influence of assuming different values of  $M_{tot}$ and $\tau$ on the PDMF of the Galaxy is shown by adopting extreme values of $M_{tot}$ (i.e., $10^9$ and $10^{14} M_\odot$) and $\tau$ (i.e., $1.55$ and $-1.2$ Gyr). Decreasing the total mass of the disk leads to a smaller present-day SFR with a deficiency of massive stars. Exponentially increasing SFRs with negative values of $\tau$ lead to a larger $b=\psi_{(t=0)}/\textless \psi\textgreater$ and to a top-heavy IMF (see also Table 1). The solid dots are data from \citet{Scalo86}. }
\label{IGIMF}
\end{center}
\end{figure*}

\begin{figure}
\begin{center}
\includegraphics[width=85mm]{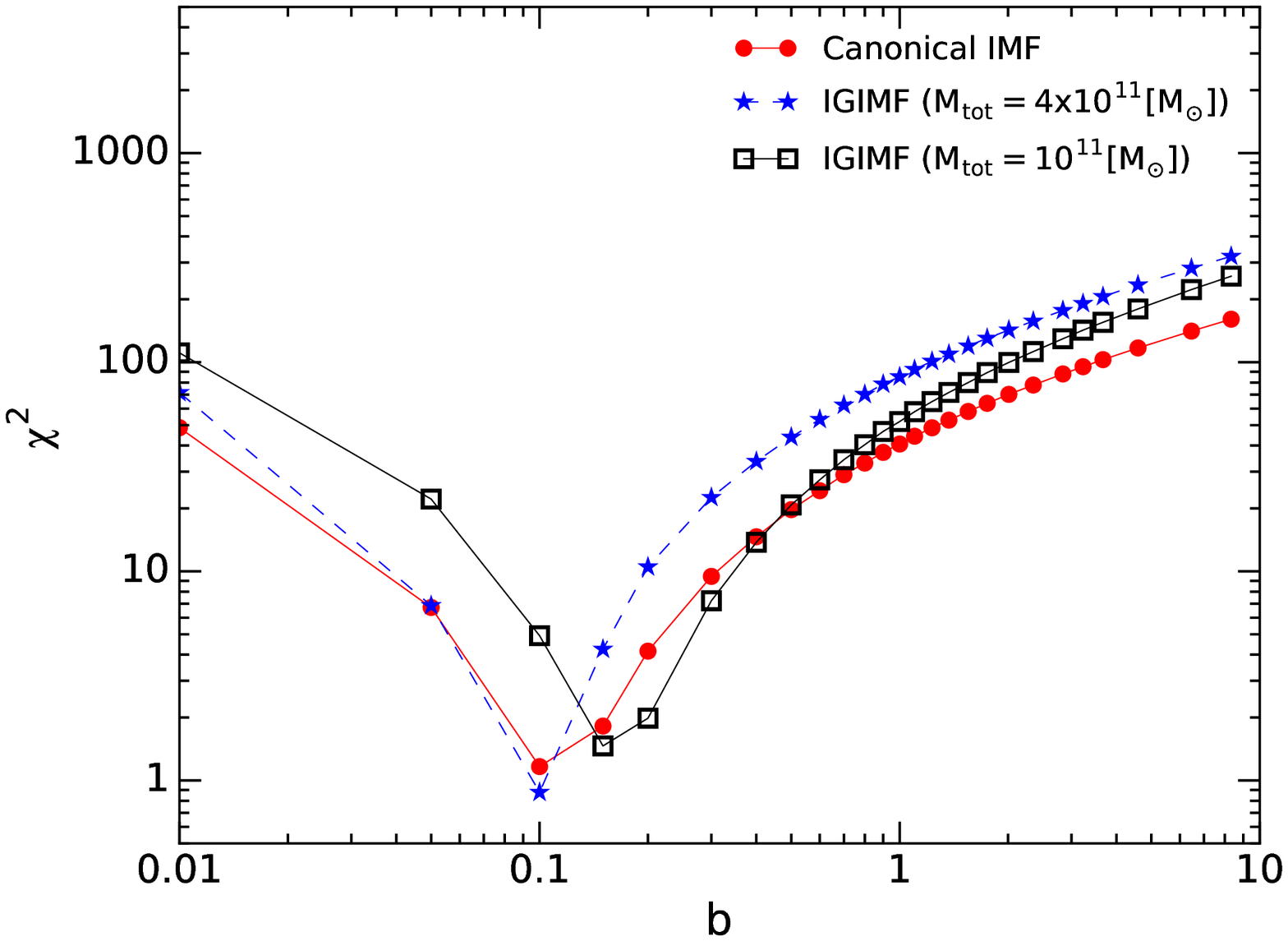}
\caption{ The $\chi^2$ values (Eq. \ref{chi2}) for the different models versus the parameter $b$ based on the IGIMF and the invariant canonical IMF.  For the IGIMF the results are plotted for two different values of the total mass:  $ M_{tot}=4 \times 10^{11} M_\odot $, which includes the best fiting model, and $M_{tot}= 10^{11} M_\odot $.}
\label{ksq}
\end{center}
\end{figure}

\subsection{Stellar Evolution}

We use the latest stellar evolution tracks from the Padova group \citep{Marigo07,Marigo08}, which exist for metallicities in the range $10^{-4} < Z < 0.030$, for ages $10^{ 6.6} < t/yr < 10^{ 10.2}$, and for initial masses $ 0.15 M_{\odot}\leqslant m \leqslant 67 M_{\odot} $. Note that we assume a time independent Solar metallicity for all stars in the Galaxy. This approximation is useful because it simplifies the calculation of the IGIMF and it is reasonable for the Milky Way thin disk population, the average metallicity of which has not evolved significantly over the past $\approx 10$ Gyr \Citep{Kobayashi11, Brusadin13}.

The contribution of stellar remnants is accounted for in the mass and mass-to-light ratio calculation.
For assigning remnant masses to dead stars we adopt the prescription of \citet{Renzini93}: stars with initial masses $m_{in} \geqslant 40 M_{\odot}$ are assumed to leave a black hole of mass $0.5 m_{in}$; for initial masses $8.5 M_{\odot}  \leqslant m_{in} \leqslant 40 M_{\odot}$, the  remnant is a $1.4 M_{\odot}$ neutron star; finally, initial masses $m_{in} \leqslant 8.5 M_{\odot} $ leave a white dwarf of mass $0.077 m_{in} + 0.48 M_{\odot}$.

Using this prescription is standard in stellar population synthesis \Citep{Maraston98,Bruzual03}. However, these relations do not take into account the dependence of remnant mass on other parameters such as metallicity.

\section{Results}\label{Sec:Results}

\subsection{Constraining the SFR from the observed PDMF of the Galaxy }

The theoretical PDMF of the Galaxy has been calculated as described in Section 2, where $\xi_{gal}=\xi_{IGIMF}$ (Eq. 2) or $\xi_{gal}=\xi_{can}$ (Eq. \ref{MF}). The total stellar mass  which is created over the age ($T_G=10$ Gyr) of the thin disk, $M_{tot}$, and the e-folding time scale parameter, $\tau $, are chosen as the two free parameters. We let the total mass of the thin disk vary in the range $\left[ 10^{9},10^{14} M_{\odot} \right ]$. Note that $M_{tot}$ is not the present-day stellar mass ($M_{*}$) of the thin disk which is composed of stars and stellar remnants.

In order to characterize exponentially decreasing and increasing SFRs, both positive and negative values, respectively, in the range $\left[ 0.1,100\right]$ and $\left[ -100,-0.1\right]$ are allowed for the $\tau $-parameter. Using these values for $\tau$ allows us to consider a constant SFR ($\tau=\infty$ ) up to extremely decreasing and increasing SFRs.  The range of adopted values for the two free parameters, $M_{tot}$, and  $\tau$ (and the corresponding $b$ value) are listed in Table 1.

For each set of the two free parameters, the corresponding PDMF is calculated.
This modeled PDMF,  $\xi_{PDMF,sim}$, is compared  with the observed Milky Way PDMF, $\xi_{PDMF,obs}$,  of main sequence stars which has been determined by \citet{Scalo86}. In Fig. 1 we show how the PDMF is sensitive to the adopted $M_{tot}$ and $\tau$ in the IGIMF theory by comparing the PDMF of the  models using two different values of $M_{tot}$ and $\tau$ with the observed PDMF of the Milky-Way. Increasing/decreasing the $\tau$-parameter in models with a decreasing/increasing SFR (which leads to a higher present-day SFR in both cases), leads the PDMF to become more top-heavy. Moreover, the present-day SFR will increase with increasing the mass of the thin disk leading to a more top-heavy PDMF.

\begin{figure*}
\begin{center}
\includegraphics[width=180mm,height=80mm]{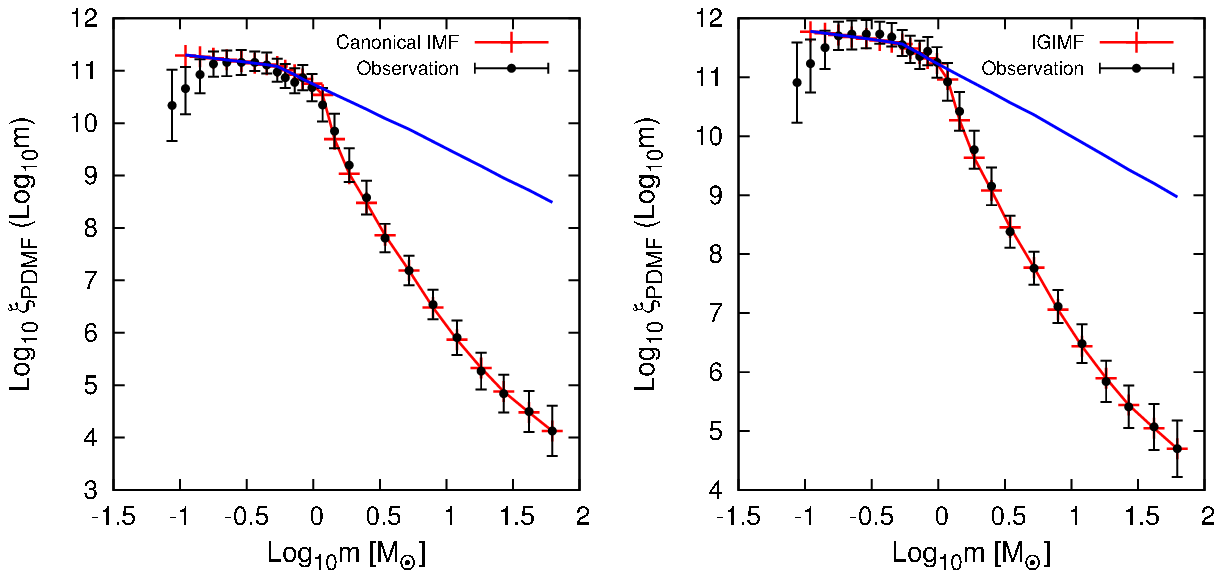}
\caption{The Milky Way disk PDMF with uncertainties derived by \citet{Scalo86} is shown as black filled circles. The best-fit models (Fig. \ref{ksq}) are shown as red crosses in both panels.  Using the invariant canonical IMF, the best-fitting model has $\tau=2.8$ Gyr and $ M_{tot}= 10^{11} M_\odot $. This implies the present-day V-band stellar mass-to-light ratio of the MW to be 1.35. In the case of the IGIMF, the best-fitting model has $\tau=2.8$ Gyr and $ M_{tot}=4 \times 10^{11} M_\odot $ and a V-band mass-to-light ratio of the MW of $2.79$. The observed MF is vertically shifted up 9.8 dex to have same number of most massive stars as the modeled PDMF (see footnote 2). In both panels the blue line shows the canonical IMF (Eq. \ref{MF}) for comparison. The decrease of the observed PDMF below $0.3 M_\odot$ stems from using an inappropriate stellar mass-luminosity relation and from not  correcting the star-counts for unresolved binary systems \citep{Kroupa93}, both not known to \Citet{Scalo86}. }
\label{fit1}
\end{center}
\end{figure*}

In order to measure the quality of the fit of the model to the observations in the mass range $m\geq1M_\odot$, we employ the $\chi^2$ goodness-of-fit test defined as:

\begin{equation} \label{chi2}
\chi^2=\sum_{i=1}^{n}\frac{(\xi_{PDMF,sim}^i(\log_{10} m_i)-\xi_{PDMF,obs}^i(\log_{10} m_i))^2}{\sigma_i^2},
\end{equation}
where, $\sigma_i^2$ is the uncertainty in the observed data. Note that the observed PDMF by \Citet{Scalo86} denotes the surface density of objects which is defined as the number of stars per pc$^{2}$ in the solar neighborhood per interval of $\log_{10} m$ (i.e., $d(N/S)/d\log_{10} m$). Therefore, the observed PDMF should be scaled by the Galactic disk mass over its surface density in the solar neighborhood to get the disk number of stars per logarithmic mass interval (i.e., $\xi_{PDMF,obs}(\log_{10} m)$), where the logarithmic PDMF is

\begin{equation}
\xi(\log_{10} m )= \frac{dN}{d\log_{10} m}=\xi(m)m\ln10. \label{IMF1}
\end{equation}

Here, we shifted up the observed surface density of the PDMF to have the same number of most massive stars at $m=60 M_\odot $ in the models as in the observed PDMF.  This is because the $60 M_\odot $ stars are the most luminous stars in Scalo's observational data set such that the uncertainty is the least and these stars are observable up to a distance of 5 kpc and hence, the obtained number density for stars with this mass can be a good representation of the global number density in the Galactic disk.  The observational data are shifted up vertically by 9.8 dex (i.e., a factor of $6\times10^9$) for IGIMF models with $M_{tot}=4\times10^{11} M_\odot$, and by 9.2 dex (i.e., a factor of $1.5\times10^9$) for models with $M_{tot}=10^{11} M_\odot$.

Note that this shift is proportional to the total mass of the Galaxy. Considering a rough approximation that the Galactic thin disc radius is about $R_{disc}=15$ kpc and its density decreases outwards exponentially with a radial scale length of $2.6 \pm 0.5$ kpc \citep{Bland16}, the solar neighborhood surface density is about $0.7$ of the average density of the thin disc such  that the observed PDMF (i.e., number of stars$/(\log m $~pc$^2$)) should be multiplied  by a factor\footnote{ $\pi R_{disc}^2\times\frac{\langle \rho \rangle_{(0 \leq  R_G \leq 15 kpc)}}{\langle \rho \rangle_{(7 \leq  R_G \leq 10 kpc)}}\approx10^9 pc^2$, assuming the PDMF as determined by Scalo (1986) samples the Galactic thin disk in the annulus  7 to 10 kpc. } of  about $10^9 $pc$^2$ to be correctly normalised to the galaxy-wide PDMF, that is to the model PDMF.
The model with a total mass of $10^{11} M_\odot$ and $b=0.2$ is in \emph{1-sigma} agreement with the observed PDMF and predicts a  total stellar mass of the Galactic thin disk, $M_*=5.8\times10^{10}M_\odot$,  and a present-day SFR, $1.5 M_\odot$ yr$^{-1}$, which is consistent with the value obtained by \cite{Licquia15} who find a SFR for the Galaxy of $\dot{M}_*=1.65\pm 0.19 M_\odot$ yr$^{-1}$ and a disk stellar mass of $5.17\pm1.11\times10^{10}M_\odot$.

The dependency of $\chi^2$  on $b$ is shown in Fig. \ref{ksq} assuming $\xi_{gal}=\xi_{can}$, the canonical IMF, and assuming $\xi_{gal}=\xi_{IGIMF}$  with two different values of $M_{tot}$.

It is necessary to distinguish between the Galaxy as a whole and the sub-population of stars nearby the Sun which is what Scalo (1986) used to constrain the Galactic-field PDMF and the Galactic-field IMF. Here we assume the star-counts Scalo (1986) used are representative of the whole  MW thin disk. That is, the field-IMF as derived from the PDMF is valid for the whole MW subject to the correct relative normalization. Thus, the mass in stars of the Galaxy is used as one constraint on the SFR and the shape of the field-IMF above $1\,M_\odot$ as the other.

We compared the PDMF of the best-fitting model based on the IGIMF and canonical IMF with the observed PDMF in Fig. \ref{fit1}. Interestingly, although we just used the observed PDMF above $1M_\odot$ to obtain the best fitting model, this model is also in good agreement with the PDMF for low mass stars.
For both canonical IMF and IGIMF, our results show an excellent reproduction of the Scalo PDMF with respectively $\chi^2=1.16$ and $0.88$, for a \emph{declining SFR} with an e-folding time scale of about $\tau=2.8$ Gyr. However, only the IGIMF model allows a prediction about the present day Galactic properties. 
The mass and light of a galaxy are calculated by an integral over the SFR,
\begin{equation}
L(t)=\int_0^{t} \psi(t-t')L'(t') dt', \label{Lt}
\end{equation}

\begin{equation}
M(t)=\int_0^{t} \psi(t-t')M'(t') dt', \label{Mt}
\end{equation}
where, $M'(t')$ and $L'(t')$ are, respectively,  the mass and luminosity of a coeval set of stars at time $t'$ after birth, and the IMF is normalized such that one solar mass of stars is created over each coeval set. Note that, using the invariant IMF, the mass-to-light ratio is independent of the total mass which is converted into stars. This is because of cancelation of the normalization parameter, C,  which is defined in $ \psi(t)$ (Eq. \ref{sfr}). In other words,  the total stellar mass acts as the normalization of the SFR (i.e., the C parameter). Therefore, it is not possible to deduce, using the shape of the observed PDMF, the total mass ($M_{tot}$), luminosity or present-day SFR in the context of the invariant IMF. In order for the present-day stellar Galactic disc mass to be comparable with the results of \citet{Licquia15, Flynn06}, we adopt $M_{tot}=10^{11}M_\odot$ for the canonical models. However, any other assumption for $M_{tot}$ will only scale the resulting present-day mass, luminosity, and SFR by the ratio of $M_{tot}/10^{11} M_\odot$. 

In the context of the IGIMF, as the slope of the IGIMF depends on the embedded cluster MF which varies with the SFR, the mass-to-light ratio varies with both the adopted $M_{tot}$ and the $\tau$ value. Therefore, the basic parameters of SFR, $M_{tot}$,  $\tau$, and luminosity can be well constrained within the context of the IGIMF theory. In Fig. \ref{IGIMF}, we compare  the PDMF  of the Milky-Way to different IGIMF models in  the two-parameter space of $M_{tot}-\tau$.

The best-fitting model using the IGIMF has been obtained for $M_{tot}=4\times10^{11} M_\odot$ and $\tau=2.8$Gyr which implies the present day star formation rate of the thin disc of the MW, and the present-day total stellar mass (including live and remnant stars) to be $\dot{M}_\ast = 4.1$ $M_\odot$ yr$^{-1}$  and $M_\ast = 2.1\times 10^{11} M_\odot$, respectively. In this model about one-third of the present day mass of the Galaxy is in the form of remnants which predicts the number of black holes and neutron stars in the MW to be about $9.8\times10^8$ and $4.7\times10^9$, respectively. The best-fitted parameters, including the present day stellar mass-to-light ratio, present-day SFR in the disk, stellar mass and number of black holes and neutron stars are listed in Table 2. The errors listed in the table are the $68\% $ confidence interval. 

The reason why the IGIMF model leads to a significantly larger number of black holes and neutron stars than the invariant canonical IMF model is that the IGIMF was top-heavy during the high SFR epoch. 

Assuming a radially exponentially declining density profile for the Galactic thin disk (with the scale length of 2.6 kpc), the surface number density of black holes, neutron stars, white dwarfs and main sequence stars as a function of Galactocentric distance are plotted in Fig. \ref{bh} for the best-fitting IGIMF model. This distribution is an important prediction from the IGIMF theory such that the observed distribution of different stellar types, can potentially be used to test this theory.
 
The declining SFR that we found here is qualitatively in agreement with the results of \citet{Bovy} and \citet{Aumer09} who report the SFR to be $2-7$ times lower now than 10 Gyr ago in the disc. However, using the best-fitting value of $\tau=2.8^{+1.7}_{-0.1}$, our resulting  SFR is faster declining such that the present-day SFR is $10-40$ times lower than its value 10 Gyr ago.
This difference could be due to the different methods and samples of data used in extracting the IMF and SFH of the MW. \citet{Bovy} anchors IMF models to the observed density of $(K2V-K4V)$ stellar type populations, because this type of stars is sufficiently  long lived to trace all mass formed over the history of the disk. However, the observed mass function of low mass stars suffers from the effect of unresolved binaries and is affected by incompleteness. Our analyses is restricted to the masses above one solar mass.  Using the results of \citet{Bovy} and \citet{Aumer09} as an additional constraint on the Galactic SFH, the best fitting model for $0.3\leq b \leq 0.7$, which corresponds to the SFR declining by $2-7$ times in 10 Gyr, is calculated. As has been shown in Fig. \ref{SFH},  adding this independent constraint, the $\chi^2$ of the best fitting IGIMF model is smaller than for the canonical IMF by a factor of 2.5. However, adopting this as a true constraint on the MW SFH, the IGIMF theory predicts a smaller mass of $3 \times 10^{10} M_\odot$, for the Galactic thin disk. The obtained Galactic properties using this constrain on the SFH are listed in Table 2. 

It should be noted that the detailed shape of the embedded cluster MF might be different from our assumption.  It is debated whether it is a power law \Citep{Whitmore07, Whitmore10, Chandar10}  or a Schechter function \Citep{Gieles06, Bastian08, Larsen09}. \citet{Lieberz17} show that the latter follows from the former  on integrating over a whole galaxy. However, because of typically the small number of high-mass star clusters the precise shape of the star cluster mass function at the high-mass end is not expected to make a huge difference in the galaxy-wide mass function \citep{Bastian08}. In order to evaluate the robustness of our results against changes in the shape of  the cluster mass function, we changed the power law index of this function  in a reasonable range. Decreasing the value of power-law index of the cluster mass function as  $\beta_1=-0.106 \log_{10} \psi(t) + 1.8$ which implies more massive clusters, the best-fitting model is obtained for $M_{tot}=0.8\times 10^{11} M_\odot$  and $\tau=2.8$. On the other hand,  adopting a steeper slope for the cluster mass function, $\beta_2=-0.106 \log_{10} \psi(t) + 2.2,$ which implies less-massive clusters, a larger MW disk mass ($M_{tot}= 8\times 10^{11} M_\odot$), and a shallower declining SFR ($ \tau=3.2$) is obtained for the best fitting model.  The results of these two models are listed in Table 2 indicated as  $IGIMF_{\beta_1}$ and $IGIMF_{\beta_2}$, respectively. As illustrated in Fig. \ref{beta}, despite the change in the $\beta-$index within a reasonable range, the results are still in agreement at  the $1\sigma$ confidence level.

\begin{figure}
\begin{center}
\includegraphics[width=85mm]{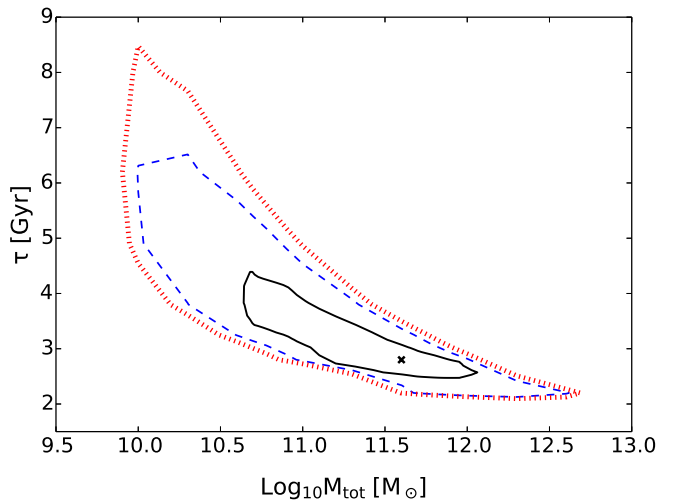}
\caption{ The 0.68\%, 0.95\% and 0.98\% confidence regions for $M_{tot}$ and $\tau$ by comparing the IGIMF model with the observed PDMF of the thin disc of the MW. See text and Table 2 for more details. }
\label{IGIMF}
\end{center}
\end{figure}

\begin{figure}
\begin{center}
\includegraphics[width=85mm]{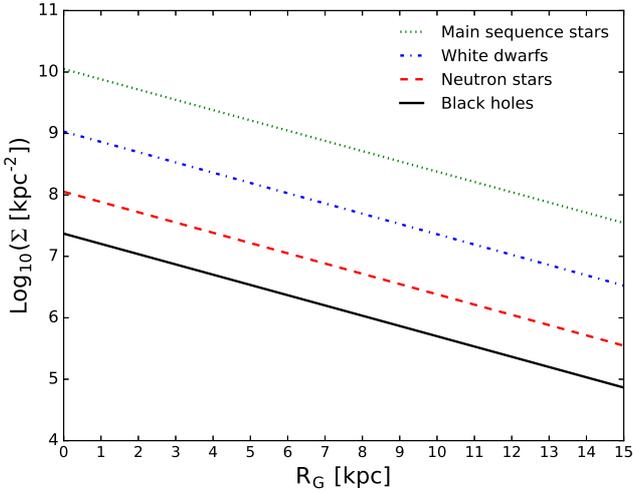}
\caption{ The number of BHs, NSs, WDs, and MS stars per kpc$^2$ vs. Galactocentric distance is plotted for the best-fitting IGIMF model. An exponentially declining profile with a radial scale length of 2.6 kpc is assumed for the mass density of the Galactic thin disk and the embedded cluster mass function assumed to not depend on $R_G$. The models investigated here assume the IGIMF to not vary radially in the Galaxy (but see \citealt{Pflamm08}).}
\label{bh}
\end{center}
\end{figure}

\begin{figure}
\begin{center}
\includegraphics[width=85mm]{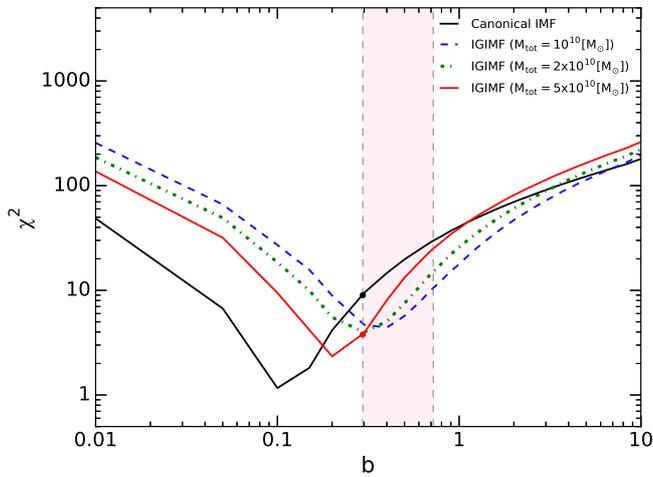}
\caption{ The $\chi^2$ values for different models versus $b-$parameter based on the IGIMF and the canonical IMF. According to \Citet{Bovy} and \citet{Aumer09}, the Galactic present-day SFR is $2-7$ times lower than 10 Gyr ago corresponding to the $b-$parameter varying in the range of $0.3-0.7$ which is shown by the vertical shadowed area. Adopting this constraint on the SFH, the best-fitting IGIMF model is obtained for $M_{tot}=5 \times 10^{10} M_\odot$ with a lower $\chi^2$ than the canonical IMF. }
\label{SFH}
\end{center}
\end{figure}

\begin{figure}
\begin{center}
\includegraphics[width=85mm]{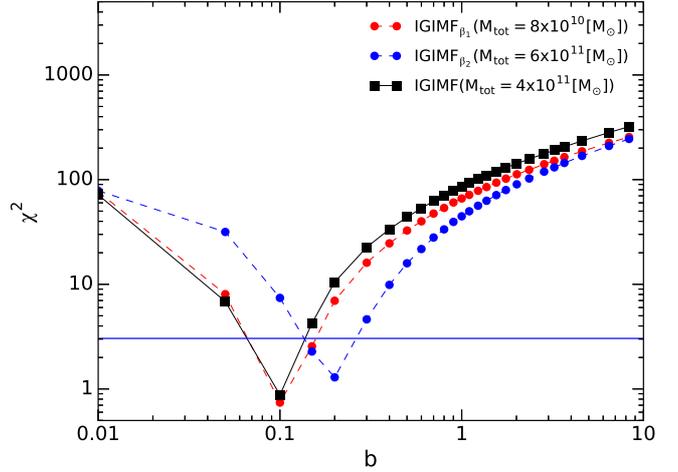}
\caption{The robustness of the best-fitting model against changes in the shape of the embedded cluster mass function. The horizontal line shows the 1$\sigma$ confidence level.}
\label{beta}
\end{center}
\end{figure}

\textbf{\subsection{Maximum cluster mass$-SFR$ relation}}

\begin{figure}
\begin{center}
\includegraphics[width=85mm]{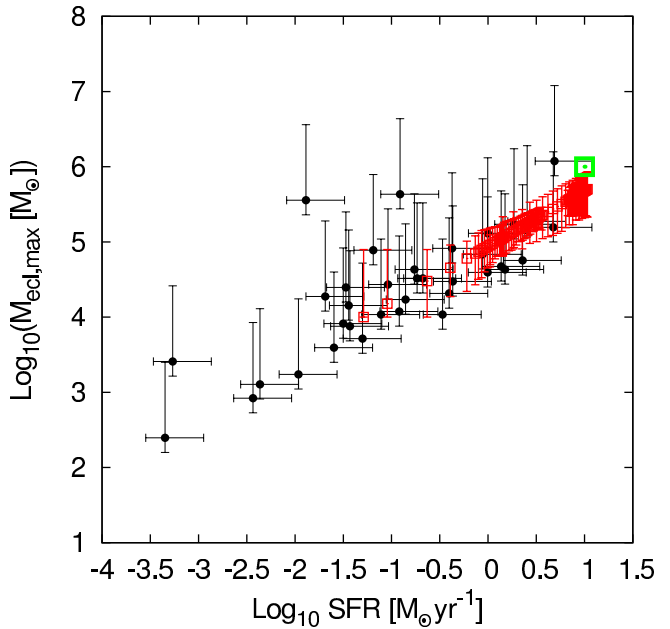}
\caption{ The dependence of the maximum embedded cluster mass on the underlying global SFR of the host galaxy, both in logarithmic units. Red open squares are the calculated $M_{ecl}^{max}$ from vertical stellar velocity dispersion data \citep{Kroupa02b}. The corresponding SFR for each $M_{ecl}^{max}$ is calculated from the IGIMF theory \citep{Weidner04} using Eq. \ref{Mtot10}. Filled dots are extragalactic observations by \citet{WKL04} based on data by \Citet{Larsen02}. The green symbol is the maximum embedded cluster mass for the thick disk which is assumed to have a mass of $M=0.2~M_{disk}$, where $M_{disk}=5\times10^{10} M_{\odot}$. We assume the thick disk formation time scale is $1$ Gyr. }
\label{ecl-SFR}
\end{center}
\end{figure}

\Citet{Larsen02} has shown that the SFR of a galaxy correlates positively with the brightest very young cluster mass observed in the galaxy. This is confirmed by \citet{Rand13} who stress that the correlation is stronger than given by random sampling from the very young (embedded) cluster mass function, confirming the conclusion by \citet{Schulz15} and \citet{Pflamm13} that the formation of star clusters is not a random process within a galaxy. \Citet{Larsen02} compiled observed luminosities of the brightest and youngest clusters in galaxies with different SFRs. These lumonosities are transformed to cluster masses by \citet{WKL04}.  In Fig. \ref{ecl-SFR} the maximum embedded cluster mass of individual clusters in the \Citet{Larsen02} data set is plotted as a function of SFR (filled dots).

The motion of stars perpendicular to the disk of the Galaxy is known to become faster for older stars. This is expressed as the vertical disk heating law. By taking into account the secular heating due to spiral density waves, molecular clouds and the mass growth of the Galaxy, and by matching the data with the  model in which all stars form in embedded clusters which are distributed as a power-law embedded cluster mass function, Kroupa (2002b) constrained the maximum embedded cluster mass which formed at a give time in the Galactic disk. In this model, clustered star formation added kinematical components to the thin disk of the Galaxy through the process of rapid gas expulsion and subsequent unbinding and expansion of a substantial fraction of the embedded cluster  \citep{Kroupa01b, Brinkmann17}. Assuming that the embedded star clusters are the building blocks of galaxies, and that the embedded cluster mass function is a power-law, \citet{Kroupa02b} found a maximum embedded cluster mass, $M_{ecl}^{max}$,  as a function of time using the age-velocity dispersion relation.
We use the resulting $M_{ecl}^{max}$'s from \citet{Kroupa02b}  and compute the corresponding SFR for each maximum cluster mass from the IGIMF theory by calculating the total mass produced in 10 Myr using Eq. \ref{Mtot10} (over-plotted in Fig. \ref{ecl-SFR} as red open squares for comparison). Interestingly, as can be seen, these calculated  $M_{ecl}^{max}$, based on the velocity dispersion of stars in the Galactic disk, are in agreement with the independent observational data by \Citet{Larsen02}. In concluding this, it should be noted that the IGIMF theory is not calibrated on the observed data by \citet{Larsen02}, instead, the IGIMF theory is based on two assumptions:
first, the mass of star clusters are distributed as a power-law function with the index $\beta$ being about 2, and second, 
the star formation time scale, $\delta t$, is about $10 Myr$ and follows from the observed offset of $CO$ arms and $H_\alpha$ arms in spiral galaxies \Citep{ Egusa04, Egusa09} which is in agreement with the observationally mapped sequence of evolution from molecular clouds to star cluster population in galaxies \Citep{Fukui99, Yamaguchi01, Tamburro08}.
 
the observational attempts to estimate the time scale of star cluster formation
\Citep{Fukui99, Yamaguchi01, Egusa04, Egusa09, Tamburro08}.

\begin{figure}
\begin{center}
\includegraphics[width=85mm]{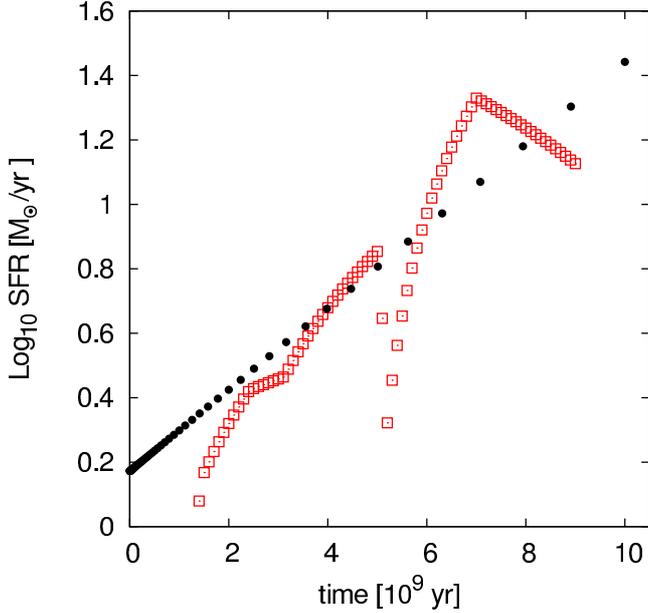}
\caption{ Red squares: the calculated SFR from vertical stellar velocity dispersion data \citep{Kroupa02b}  which is shifted up by a factor of 2.3 (see text for more details). Black circles: the predicted SFH for the MW from the best-fitting PDMF based on the IGIMF model with  $M_{tot}=10^{11} M_\odot$ (Table 2).The present time is at t=0 }
\label{ecl-sfr2}
\end{center}
\end{figure}

We also compared  these calculated SFR's for each maximum cluster mass, with the derived star formation history (SFH) of the MW  based on the IGIMF model with $M_{tot}=10^{11} M_\odot$ (Fig. \ref{ecl-sfr2}). Note that the velocity dispersion is measured in the Solar neighborhood where the SFR and the local maximum cluster mass is lower than the global value. The integral over the calculated SFRs from the observed velocity dispersion results in $ M_{tot}=4.5\times 10^{10}$.
Therefore, the calculated SFRs in Fig. \ref{ecl-sfr2} are scaled by a factor of 2.3 to the level of the SFH of the MW  based on the IGIMF model with $M_{tot}=10^{11} M_\odot$.

Many disk galaxies at high redshift (i.e., $z>1$) have peculiar morphologies dominated by several massive giant clumps aligned within an underlying disk, appearing as "chain galaxies" when observed edge-on \Citep{vandenBergh96, Elmegreen04}.
According to Fig. \ref{ecl-SFR},  the MW appears to have looked like any disk galaxy following the observed trend of $M_{ecl}^{max}$ with SFR. From Fig. \ref{ecl-sfr2} it is seen that at high redshift the MW had a larger SFR, about $20 M_\odot/yr$  10 Gyr ago, which produced massive clusters. This SFR implies the most massive cluster to be $M_{ecl,max}=3 \times 10^7$ and about 10 clusters above $10^6 M_\odot$  will have been produced in that time. The small number of these clusters and their high luminosity increased by the top heavy IMF \citep{Marks12} lets them be detectable clearly as a few individual clusters. The MW would thus have appeared as a chain galaxy.\\

\section{Conclusion}

We here address the shape of the PDMF by applying the canonical IMF and the IGIMF theory with a time-dependent SFR. The shape of the Galactic-field PDMF for $m>1\,M_\odot$ {\it and} the normalization of the Galactic field PDMF for low mass stars allow us to constrain the star-formation history of the Galaxy and thus its present-day SFR. This then also provides us with the total mass in stars of the Galactic thin disk.

We calculated a set of models to study the influence of two basic parameters of the star formation history of a galaxy, i.e., the e-folding time scale ($\tau$) and the total disk mass ($M_{tot}$), on the present-day properties of the Galactic thin disk. Accordingly, we constrain the MW thin disk properties including its present-day SFR, mass-to-light ratio and stellar mass.  Our main conclusions are the following:

\begin{enumerate}
\item Both, the IGIMF and the canonical IMF reproduce the Galactic PDMF very well and equivalently. However, if we constrain the SFR to decline in time by a factor of 2 to 7 \citep{Bovy, Aumer09}, the degeneracy breaks and the IGIMF model is more successful in reproduction of the MW PDMF.    
\item Assuming the canonical IMF and the IGIMF, the constructed models with $\tau=2.8^{+1.7}_{-0.1}$ Gyr give the best fit. This implies that the Milky Way SFR is an exponentially declining function.  That the IMF and IGIMF models give very similar $\tau$ value  is because the MW is an intermediate-mass galaxy in which the IGIMF is similar to the canonical IMF, while the IGIMF is significantly different in  dwarf galaxies with low SFRs or massive galaxies with high SFRs \citep{Yan2017}.
\item Since the IGIMF depends on the SFR and metallicity, the IGIMF theory allows the IGIMF-slope to depend on the total mass of the host  galaxy. Hence, the Galactic present-day mass and SFR can be constrained based on the IGIMF theory, which is not possible in the context of an invariant IMF. Based on stellar population synthesis and using the canonical  IMF as the galaxy-wide IMF, we conclude that the V-band $M_\ast/L$ ratio of the Galaxy is  $1.35^{+0.06}_{-0.17}M_\odot/L_\odot$. The  IGIMF theory gives $M_\ast/L_V=2.79^{+0.48}_{-0.38}M_\odot/L_\odot $. Also in the context of the IGIMF, we obtain that the present-day stellar mass of the MW is about $M_\ast=2.1\times10^{11}M_\odot$ including $1.4\times10^{11}M_\odot$ of  living stars and $0.7\times10^{11}M_\odot$ of remnants which predicts the number of black holes and neutron stars in the MW to be about $9.8\times10^8$ and $4.7\times10^9$, respectively. The IGIMF theory predicts a larger number of BHs and NSs in the disk compared to the canonical IMF.   From the best-fitting model in the  IGIMF theory,  we find a present-day SFR of $\dot{M}=4.10^{+3.10}_{-2.80} M_\odot yr^{-1}$ which is consistent with the recent estimates of the Milky Way's SFR \citep{Bovy,Licquia15}. For the constrained SFH models $M_\ast/L_V=1.87^{+0.85}_{-0.03}M_\odot/L_\odot $, $M_\ast=3.0\times10^{10}M_\odot$ and $\tau=4.9^{+.7}_{-1.1}$ are obtained (Table 2). 

\item Since the MW would have had a higher SFR about 10 Gyr ago in the IGIMF theory it would have been populated by embedded clusters extending in mass up to a few$\times 10^6 M_{\odot}$ (Fig. \ref{ecl-SFR}). The MW could have thus appeared like a chain galaxy.

\end{enumerate}

\bsp \label{lastpage} \end{document}